# Assessing Learning in Small Sized Physics Courses

Emanuela Ene[1],
*Department of Physics & Astronomy, Texas A&M University, College Station, TX 77843-4242*
Bruce J. Ackerson[2],
*Department of Physics, Oklahoma State University, Stillwater, OK 74078-3072*

**Abstract**

We describe the construction, validation and testing of a concept inventory for an *Introduction to Physics of Semiconductors* course offered by the department of physics for undergraduate engineering students. By design, this inventory addresses both content knowledge and ability to interpret content via different cognitive processes described in Bloom's taxonomy. The primary challenge comes from the low number of test takers. Since *the Rasch Model (aka 1PL IRT model)*, can be used with small sample sizes, we describe *Rasch Modeling* analysis and results for this concept inventory.

## I. Introduction

The concept examination, described here, grew from a need to assess the effectiveness of a new pedagogy for a junior level course PHYS3313 at Oklahoma State University (OSU). Originally, the content was of a standard introductory modern physics offered in a one-semester course. Since the enrollment evolved to include mainly students in electrical and computer engineering, the departments of physics and electrical engineering collaborated to change the emphasis to an introduction to solid-state devices with a non-traditional pedagogy described by Cheville et al.[1] (Currently, as different instructors are assigned to teach this class, the pedagogy shifts between traditional and non-traditional.) The work on calibrating a concept inventory grew from a desire to assess learning for the different pedagogies.

The most popular assessment instruments are the concept inventories, starting with the Force Concept Inventory (FCI) that discriminates Newtonian from Non-Newtonian thinkers [2,3]. Today there are a plethora of concept inventories for a variety of disciplines[4]. The effort to develop such assessment instruments has helped identify what are the most important concepts and cognitive skills targeted by similar courses taught by different instructors. Although there are diagnostic instruments calibrated with thousands of test takers, there are only few assessment instruments for specialty courses. The greatest difficulty in assessing specialty courses comes from the fact that they enroll relatively small numbers of students.

Knowledge and knowledge interpretation, the two constructs that are to be measured with a concept inventory, are defined in this manuscript in the light of Reif's model[5]. Conceptual knowledge is the kind of knowledge that may be transferred between situations and can be used for inferences[6,7]; it can be achieved through reflective learning. According to Reif, experts "often use several types of knowledge redundantly to look at the same situation from different points of view" while a typical novice student has a faulty knowledge "predominantly because of a failure to heed appropriate applicability conditions."

We devised a modality to measure the coherence of knowledge, which characterizes "deep understanding." Coherence of knowledge ensures consistency, facilitates debugging, and helps regenerate knowledge that is partially forgotten. Confidence in one's knowledge is a requirement for applying available knowledge. The cognitive levels involved in operating with conceptual knowledge range from understanding to evaluation [8].

The probabilistic model for knowledge tests, first introduced by Georg Rasch [9,10] and sometime called one-parameter IRT logistic model [11], separates the calibration of the items in the test from the population distribution. "…we have no reason for considering a normal distribution of test scores as evidence for equality of units" says Rasch[12] when comparing the classical test theory and his model. A test with the items calibrated to be invariant across various samples of the population has the specific objectivity of a yardstick, as opposed to a calibration assuming a normal distribution of the scores: "When a man says he is in the 90th percentile in math ability, we need to know in what group and what test, before we can make any sense of his statement. But when he says he is 5' 11" tall, we don't ask to see the yardstick."[13] Consequently, the Rasch model increasingly finds use in concept inventory analyses [14] and some authors [15] even plead for its use as opposed to classical test theory.

---

[1] Electronic mail at ene@physics.tamu.edu
[2] Electronic mail at bruce.ackerson@okstate.edu

One of the benefits of the Rasch modeling is that it does not require large sample size. One recent concept inventory study used Rasch modeling to analyze a 24-item test, the Relativity Concept Inventory (RCI)[16] . As opposed to other concept inventories, RCI was calibrated on a small sample size: 70 persons at pre-test, 63 at post-test, out of which 53 matched repeated measures. One of the foundational books in Rasch modeling, "Best Test Design"[13], is based on the analysis of a sample of 35 children and 18 dichotomous items. According to Linacre, a world leader of Rasch measurement research, the minimum sample size for calibration of an objective scale is related to the desired range of stability of the dichotomous items[17] . The recommended minimum sample size range for a 95% confidence is 30 for ±1 logit stability and 100 for ±1/2 logit.[3] For definitive or high stakes tests of 99% confidence, Linacre recommends a sample size of 20 L, where L is the test length in items. "As a rule of thumb, at least 8 correct responses and 8 incorrect responses are needed for reasonable confidence that an item calibration is within 1 logit of a stable value."[18] On the other hand, there are issues that arise when test calibration uses too large a sample population[19,20]. The discussion pertains to goodness of fit and Type I or Type II errors in hypothesis testing. This raises the well-known statistical phenomenon of significance but not relevance in hypothesis testing.

The Rasch model assumes that answering an item in a knowledge test is a stochastic process similar to hurdling [21] and that the probability to answer an item depends on two unknown parameters: the ability of each person and the difficulty of each test item. In this paper we will adopt the symbols $\theta_v$ for person's parameter ability and $\beta_i$ for item's parameter difficulty (height of the hurdle). Assuming statistical independence of the items and of the answers, the probability for person v to answer a dichotomous item i is given by

$$P[i, v] = \frac{e^{x_{i,v}(\theta_v - \beta_i)}}{1 + e^{(\theta_v - \beta_i)}}$$

where $x_{i,v}$ has value 0, if the answer is incorrect, and 1 if correct[22]. So, when $x_{i,v} = 1$ we have the probability that the person v answers question i correctly. Fisher [23] demonstrates that such a model may be calibrated to produce an objective scale, which orders the difficulty of the items independent of the test takers. Likewise, because of the symmetry of the probability function, the population of test takers becomes ordered according to ability. The item parameter values should be invariant for different sample splits within the population or to a change of populations. To place the item parameters on an interval scale requires only the "sufficient statistics" of the raw test score for each person (a sum over item scores for each person) [24].

The parameters $\theta_v$ and $\beta_i$ are unknown quantities to be estimated by a conditional maximum likelihood fit. Likelihood is the probability that a certain collection of outputs {$x_{i,v}$} is obtained from a set of input parameters[25] (e.g. person and item scores). Holland[26] showed how to recast the likelihood function in terms of the item parameter only, the conditional maximum likelihood (CML) method. If the likelihood has a maximum as a function of the parameters {$\hat{\beta}_i$} given the known collection {$x_{i,v}$}, these are the most likely values (expected values) for the item parameter. The astute reader will recognize the correspondence with maximizing the entropy in statistical mechanics. Rasch determined maximization parameters by tedious hand calculations. When computers became available, computer programs determined parameter values using the best approximations at the time[27,28]. The open-source eRm algorithm[29] uses the estimated invariant parameters {$\hat{\beta}_i$} to find the person ability values.

A Rasch Modeled (RM) instrument ranks both the item difficulty and the person ability on a common logit scale. Zero ability does not imply the absence of ability for a person but only a fixed point on the continuum of the ability scale. If the ability is estimated based on a Rasch-scalable set of items, the zero on the ability scale is usually chosen at the median of the estimated item difficulties. For a person of zero ability, the odds of answering correctly rather than incorrectly to an item of difficulty $\beta_i$ equals $e^{\beta_i}$. The 0-logit ordinate of the RM fitting line corresponds to a 50% probability of a correct answer, when the person's parameter $\theta_v$ equals the magnitude of the item difficulty.

---

[3] 1-logit is the unit of measure for the parameters estimated via logistic models. 1-logit signifies that the natural logarithm of the odds of an event equals one. This is equivalent to a probability of $\frac{e}{1+e}$ =73% to observe the event.

Due to the specific objectivity of the Rasch calibration, the person estimated ability does not depend on the specific items chosen from a large calibrated item pool. From both theoretical and practical perspective, the requirement for representativeness of the sample is obsolete in terms of a true random selection process[30]. Because a CML Rasch calibration uncouples student ability from item difficulty, a test bank may be developed and expanded with items of known content and difficulty. As all the items in a Rasch calibrated test have equal discrimination, it does not matter what items are selected for estimating the ability of the respondents. Items may then be selected to align better with the person ability in that sample. New items can be developed to separate more finely the abilities of a target group. Or, if done electronically, a fine estimation may be done for each individual student by successively narrowing the pool of items. The real issue becomes starting the process individually and collaboratively making an item pool to which others may add.

## II  Test Development & Methodology

### A.  Selection of content area

The Physics of Semiconductors Concept Inventory (PSCI) instrument was built as a criterion-referenced test. The instrument intends to measure the ability of a person to operate with principles, laws, and idealized theoretical models related to the physics of semiconductors. Ten concepts categories identified as important for the physics of semiconductors were initially selected for item writing. Twelve experts from seven universities provided their choice of the most important concepts for the physics of semiconductors. An additional list of the ten most important concepts was built by one of the authors based on a survey of syllabi and text-books currently in use for introductory courses of the physics of semiconductors at US universities[4]. A frequency chart of the concept entries was built and the most frequent cited entries were retained.

### B.  Item writing

Item stems were written both in question format and as open-end sentences to be completed by one of the options. One third of the stems contain visual stimuli: photographs, diagrams and graphs. One quarter of the items have options in graphical format. Items were labeled with a unique identifier giving the concept category and the position in the sequence. For example, item **8.2** pertains to the concept **8**-Statistics and is the second in the sequence of the items written for this concept.

The four-option dichotomous items were written for measuring a single component of the knowledge dimension, conceptual knowledge, and four components of the cognitive dimension: understanding, application, analysis, and evaluation. A list of characteristic operations for each of the cognitive processes in Bloom's taxonomy was created and employed for item writing. Each item was assigned to the highest level of the cognitive process necessary to reach the correct answer. We exemplify with two items written for the PSCI, **8.2** at Application level and **7.2** at Evaluation level.

---

[4] The text-books in the concepts survey are listed in the *Appendix*.

*Item 8.2)*

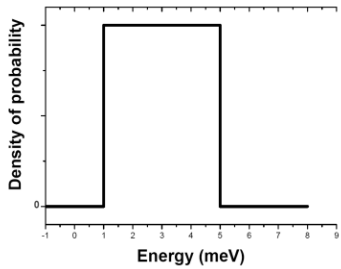

The figure above shows a hypothetical probability distribution function versus energy for a system of electrons at thermal equilibrium. The plot extends both to the left and to the right with no change in probability.
For this system, the probability that a state having the energy between 4meV and 5meV is empty is
        A. 0 %
        B. 25 %
        C. 50 %
        D. 75 %

*Item 7.2)*
An engineer tests the photoemission of a material XXX. Photoemission means the interaction of light with a material that can produce an electric current measured using an ammeter. The engineer mounts an ideal ammeter, able to detect even the tiniest electric current, as shown in the figure below.

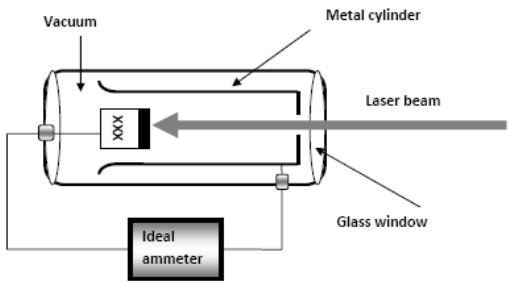

The material XXX has equal sensitivity for blue and for green light. Blue light has higher frequency than green light. When the material XXX is illuminated with a 1mW green laser beam, the ammeter reading is 0.01mA. Which of the following statements best describes the ammeter response when the material XXX is illuminated with a 1mW blue laser beam?
        A. No change in the reading.
        B. A zero value ammeter reading.
        C. A non-zero value smaller than for the green laser reading.
        D. A non-zero value greater than for the green laser reading.

For item **8.2**[5], interviews revealed that persons in field testing had difficulties interpreting an empty state as the complement event for "occupied state", at Remembering-level. The missing factual knowledge made this item written for Application-level more difficult that other items written for higher cognitive levels. A difficult step for solving the item **7.2**[6] was the connection at Analyze-level between the pieces of information "same 1mW" and "blue light has higher frequency than green light". As revealed by interviews, some solvers interpreted "1mW" as a measure of the photon number; this was a mistake at understanding-level, reflecting missing factual knowledge (knowledge of the units of measure). When answering, mistakes can be generated by missing pieces of knowledge and by the lack of training with organizing and matching pieces of information. We hypothesized that the attribute "item difficulty" reflects the amount of expert quality knowledge addressed and the level of the cognitive processes involved in answering that item. Our hypothesis was consistent with Reif's statement that experts employ several types of knowledge redundantly as opposed to the novices who lack skills for error correction.

---

[5] The suggested solving sequence for the item **8.2.** is: **Remember** (that the probability of occurrence of a range of events represents the area under the probability density function, that the total area equals 100% for a certain event, and that an empty state represents the complement event or the no-occurrence), **Understand** (that only the area from 1meV to 5meV contributes to the total probability, that probability is equally distributed on energy intervals, and that the probability of an energy interval equals the percentage of the corresponding area), **Apply** (the definition of the probability of an energy interval to calculate a 25% chance that the interval from 4meV to 5meV is occupied), then A**pply** (the definition of the complement event to calculate the chance that a state in that interval is empty as p=100% - 25%=75%).

[6] The suggested solving sequence for the item **7.2** is: **Remember** ( the equation giving the photon energy as a function of frequency, the definitions of average power and average electric current, the mechanism of the photoelectric effect, and the fact that electrons carry electric charge), **Understand** ( that
a green photon has enough energy to release an electron from the material XXX, that there is an equal probability to release an electron by either a green or a blue photon, that the 1mW value means that the two laser beam have equal power, and that the total charge generated is directly proportional to the number of photoelectrons), **Apply** ( the photon energy formula to conclude that blue photons of higher frequency have greater energy than green photons; the definition of the photoelectric effect to conclude that the rate of the emitted electrons equals the rate of incident photons; the information about equal probability of emission for green and blue to conclude that the rate of the incident photons determines uniquely the rate of emitted electrons; the average current formula to conclude that the ammeter reading is directly proportional to the number of electrons per unit time), **Analyze** the cause and effect implications ( **because** constant sensitivity of the photomaterial, at equal photon rates there will be equal ammeter readings; **because** equal incident power, the photon rate for blue is lower than the photon rate for blue), and finally **Evaluate** that a lower photon rate for blue implies a smaller ammeter reading for blue than for green.

*Table 1. Chronological development of the PSCI and schedule for test administration*

| PSCI test version | Number of items | Date | Persons* | Comment |
|---|---|---|---|---|
| Open-ended questions deployed | >100 | Aug 2009 | 72 | Experts and novices |
| PILOT 1 | 47 | Dec 2010 | 14 | post-test |
| Interviews conducted | | Dec 2010 | 5 | |
| PILOT 2 | 45 | Jan 2011 | 5 | pre-test |
| PILOT 2 | 45 | April 2011 | 11 | post-test |
| Interviews conducted | | Dec 2010 | 8 | |
| PILOT 2 | 45 | Aug 2011 | 12 | pre-test |
| PILOT 2 | 45 | Dec 2011 | 7 | post-test |
| Interviews conducted | | Dec 2011 | 1 | |
| ALPHA | 30 | Jan 2012 | 10 | pre-test |
| ALPHA | 30 | Apr 2012 | 24 | post-test |
| BETA | 18 | Apr 2013 | 35 | post-test, independent sample |

Note. *Only the persons answering all the items in the PSCI test were counted

### C. Participant selection

The targeted population was "undergraduate students who are or have been enrolled in a physics of semiconductors or semiconductor devices introductory course at a university using English language for instruction." There were two categories of participants involved in the PSCI development. Undergraduate and graduate students were called "novices." University professors teaching semiconductors courses or conducting research in the field were called "experts." Participants enrolled on a volunteer basis.

Experts were selected by snowball sampling by sending invitational e-mails to more than one hundred faculty in the field of semiconductors. The group of novices was recruited by e-mailing to 248 undergraduate students enrolled in the PHYS3313 class between Fall2008 and Spring2012 and to 25 graduate students from the Department of Physics at OSU that had not taken a semiconductors' course. To assist the reader, **Table 1** presents details of the chronological development of the concept inventory.

### D. Field testing

Field testing was performed at Oklahoma State University (OSU) between December 2010 and April 2012. There were 66 undergraduate and 2 graduate students in field testing. At the beginning of each test administration, the novice participants were advised to answer according to their best knowledge to all the items. They were told that there is no right or wrong answer, as the PSCI test is under development.

In all the stages of the field-testing, the PSCI was offered as an online quiz, available via a lock-down browser, in the proctored environment of a computer lab. Backward navigation through pages was prevented and no feedback for either correctness of the answers or total score was provided after test administration. A single attempt was allowed with the completion time restricted to 100 minutes. All the participants finished before the time limit.

Thirty items (ALPHA version) were present with insignificant variations of wording or of the order of the options in all the versions of the PSCI test. Rasch analysis performed for a pooled sample of 83 response patterns showed that these 30 items satisfy the Rasch model.

### E. Validity of content and construct

Validity certifies whether or not the test measures what it says it measures, as well as what interpretations can be inferred from the scores on the test. According to Crocker and Algina[31], the three major types of validation studies are: content validation, for the situations when the test scores will be used to draw an inference to a larger domain of items similar to those in the test itself; criterion-related validation when the test scores are the base for inferences about the performance on some real behavioral variable of practical importance; construct validation when the quality to be measured is defined by the test itself.

Evidence of content-related validity was established early in the PSCI test development process. The initial pool of items was rated by six different judges (raters) for relevancy of the concepts tested using a five point Likert scale. The raters also were asked to use a three point Likert scale for the clarity of each of the items. Only 47 items with positive ratings were deployed for field testing.

The evidence of criterion-related validity was established after the pilot study by analyzing the interviews with the participants. The good fit on the Rasch linear scale and the confirmation of the invariance of calibration on an independent sample of persons in April 2013 provided evidence for construct-related validity.

## III. Data Analysis

### A. Classical analysis

The information collected from the PSCI field-testing was separated into item data and persons' characteristic data. The item data were examined for distractor effectiveness, for item functioning and for selecting a minimal set of items for future modeling. The answering patterns for each person (person data) were examined for identifying if the sample in field testing was homogeneous or not.

Classical item analysis, which is not shown here, revealed that the items of the PSCI test have a major underlying factor that could explain the data variance. A coefficient Guttmann Lambda 6[32], greater than 0.8, implied that the test is reliable in classical interpretation. The raw score distribution was not normal for the subsamples in field test.

We investigated the difference between the students taking the PSCI test before instruction and those tested after the instruction in PHYS3313 by principal component VARIMAX analysis for detecting the influence of the variable "instruction in PHYS3313". No significant sample split due to instruction was found. Linear discriminant analysis LDA was employed for identifying possible classes due to factors such as "time since taking the course", gender, first language, or GPA. No significant difference due to instruction, time since the course was taken, or to the demographics was detected by LDA. The classical approach found no significant class split for the answering patterns collected during field-testing. We concluded that we had just one population in field testing.

### B. <u>Rasch Modeling Results</u>

#### 1. **Goodness-of-fit and item easiness estimation**

The eRm[29,33] R-package was employed to fit the 30 items of PSCI-ALPHA on a Rasch linear scale. The linear calibration of the PSCI-ALPHA, independent of the person distribution, is shown in **Figure 1**. The good fit of the 30 items on the Rasch one-dimensional scale supports the idea that there is a unique construct measured by the 30 items. We labeled the construct measured by PSCI-ALPHA as "composite-ability to operate with conceptual knowledge of physics of semiconductors". The easiest item is **7.1** with an estimate of β[7.1]= -1.31. The most difficult is **4.4** with an estimate of β[4.4]= 1.24.

The estimated beta-parameters reflect in principal the knowledge content and the cognitive processes implied in answering each item. It may occur that an item attributed to the superior cognitive level Evaluation, like the item **1.2**, is less difficult (has a smaller beta-parameter) than an item attributed to the lower cognitive level Application, like the item **8.2**. As we will show, Rasch scaling allowed us to separate the items in subscales and to track directly knowledge and knowledge coherence.

The graphical check for the likelihood ratio goodness-of-fit [34] showed that the responses to the items in the PSCI-ALPHA test are invariant across distinct subgroups in field testing. The group split criterion in **Figure 2** is the median raw score. On the vertical axis are the estimated beta-parameters obtained by maximizing the likelihood for the group of persons with raw scores greater than the median while on the horizontal axis are the beta-parameters estimated by maximizing the likelihood for the group of persons with raw scores lower than the median. The unit on both axes is logit. The estimated beta-parameters characterize the group's thresholds. The 95% confidence interval of each estimated threshold is plotted as an ellipse[35]. Items that have equal estimated beta-parameters for the two split groups lie on the diagonal line in **Figure 2**. It appears that the items are group invariant in the limit of the error of measurement. The extreme items 7.1 and 4.4 are the furthest from the item functioning invariance line.

The informal parameters Outfit Mean Squares and Infit Mean Squares[36] were employed for checking the goodness of fit (Gof) of the model. The Outfit Mean Square statistic, which measures the average mismatch between data and model, was useful to identify outliers in the sample of persons in field testing. By cross-referencing with the answering patterns, it was found that the outlier persons did not answer all the 30 items. The data from the outliers were discarded and the field test sample was reduced to 83 response patterns. The Infit Mean Square and the approximate t-statistic[37] Infit-t helped identify those items generating outlier and bias. The usual requirement for an unbiased Rasch scale is that |Infit-t| ≤ 2. The results of the Rasch calibration are shown in **Table 2**. One should note that only one item in PSCI-ALPHA, **7.1**, had an Infit-t statistic greater than 2.

2. **Person ability estimation with the 30 item PSCI-ALPHA**

A Conditional Maximum Likelihood CML algorithm estimates the item parameters independent of the persons taking the test. The estimated item parameters are substituted back into the algorithm to estimate person abilities, which are function of only the raw score. For the PSCI-ALPHA range of scores, the person composite-abilities range from almost -2logit to 0.43logit. This spread is due to the variability of the sample used for field-testing. According to Linacre[17] an 1-logit size interval gives a good person location for most of the practical applications. The Rasch calibrated PSCI-ALPHA instrument locates a person within 0.802-logit average sized 95% confidence interval. The standard error calculated for the sample in field testing is consistent with the theoretical average value of 0.4472logit for a 30 item instrument [38]. The standard error of measurement (SEM)[39] varies across items and person samples.

The score-ability conversion function can predict person parameters for scores not yet observed; this represents a major benefit of the Rasch calibration. Therefore the Rasch calibrated PSCI-ALPHA test can be employed to measure the basic knowledge of physics of semiconductors across universities, for groups of persons with a lower or higher ability than the sample in field-test.

In **Figure 3**, the item parameters and the person parameters are plotted on the same scale. The common scale origin is placed at the item of median difficulty. It should be noted that PSCI-ALPHA is not centered on the abilities of the sample in field test: the median of the person composite-ability lies below the median item-difficulty. PSCI-ALPHA is most accurate for composite-abilities close to 2 logit, where the error of estimation of the item-scale has a minimum.

3. **Measuring the deep understanding of the concepts**

A minimal set of 18 items, all with the Infit-t statistic below 2, was selected as the PSCI-BETA instrument. The items in PSCI-BETA are dedicated to six particular concepts: carrier transport, energy bands, carrier concentration, p-n junction IV plot, quantum approach, and statistics. Answering these items involves recalling and understanding the concepts and the laws describing them but Application is the lowest level of the cognitive process in Bloom's taxonomy tested in order to answer an item without guessing. The concept map for the PSCI-BETA is presented in **Table 3**.

The 18-item scale was further split into three separate subscales each meeting the invariance requirement, one for each cognitive level Application, Analysis, and Evaluation. A fourth subscale labeled Pairs consist of six pairs of correlated items that were fit with the Partial Credit Model[40] and constitute a linkage between the other three subsets. The score for the subset Pairs provides extra information about the responses consistency and a measure of the construct "deep understanding." The internal structure of the PSCI-BETA instrument is pictured in **Figure 4**.

The construct "deep understanding" measured by the PSCI-BETA with its subscales has two aspects: quality of the conceptual knowledge and consistency of the knowledge. The quality of the conceptual knowledge is measured by the partial scores for the items at Application-, Analysis-, and Evaluation-level. In particular, the ability to infer new knowledge from existing knowledge is needed for answering the items at evaluation level. The consistency of the knowledge is measured by the partial score for answering pairs of items addressing the same concept presented in different instances.

*Table 2: Item fit statistics for PSCI-ALPHA*

| Item | Outfit MSQ | Infit MSQ | Outfit t | Infit t | Item | Outfit MSQ | Infit MSQ | Outfit t | Infit t |
|---|---|---|---|---|---|---|---|---|---|
| **1.2** | 1.047 | 1.041 | 0.96 | 0.92 | **5.3** | 1.014 | 0.973 | 0.18 | -0.2 |
| **2.2** | 1.079 | 1.064 | 1.99 | 1.74 | **5.5** | 0.955 | 0.965 | -0.93 | -0.8 |
| **2.3** | 1.075 | 1.067 | 2.01 | 1.94 | **5.6** | 0.915 | 0.919 | -1.27 | -1.42 |
| **2.4** | 0.949 | 0.959 | -0.69 | -0.65 | **6.1** | 1.09 | 1.066 | 1.93 | 1.65 |
| **2.6** | 0.924 | 0.939 | -1.61 | -1.42 | **6.2** | 0.961 | 0.959 | -0.81 | -0.96 |
| **3.2** | 1.055 | 1.05 | 0.63 | 0.67 | **6.4** | 0.98 | 0.993 | -0.22 | -0.07 |
| **4.1** | 0.948 | 0.953 | -1.35 | -1.34 | **7.1** | 0.799 | 0.842 | -3.06 | -2.86 |
| **4.2** | 1.047 | 1.033 | 0.66 | 0.55 | **7.2** | 1.171 | 1.065 | 1.14 | 0.55 |
| **4.3** | 1.033 | 1.023 | 0.72 | 0.56 | **7.4** | 0.946 | 0.953 | -1.07 | -1.05 |
| **4.4** | 0.849 | 0.919 | -0.86 | -0.52 | **7.5** | 1.007 | 1.005 | 0.19 | 0.17 |
| **4.6** | 0.967 | 0.96 | -0.64 | -0.88 | **7.6** | 1.054 | 1.041 | 0.65 | 0.58 |
| **4.7** | 0.981 | 0.983 | -0.39 | -0.4 | **7.7** | 1.01 | 1 | 0.19 | 0.03 |
| **4.8** | 1.133 | 1.073 | 1.67 | 1.11 | **8.2** | 0.967 | 0.972 | -0.29 | -0.29 |
| **5.1** | 1.129 | 1.089 | 1.54 | 1.26 | **8.3** | 1.026 | 1.024 | 0.44 | 0.46 |
| **5.2** | 0.937 | 0.932 | -1.12 | -1.38 | **10.1** | 0.966 | 0.967 | -0.87 | -0.92 |

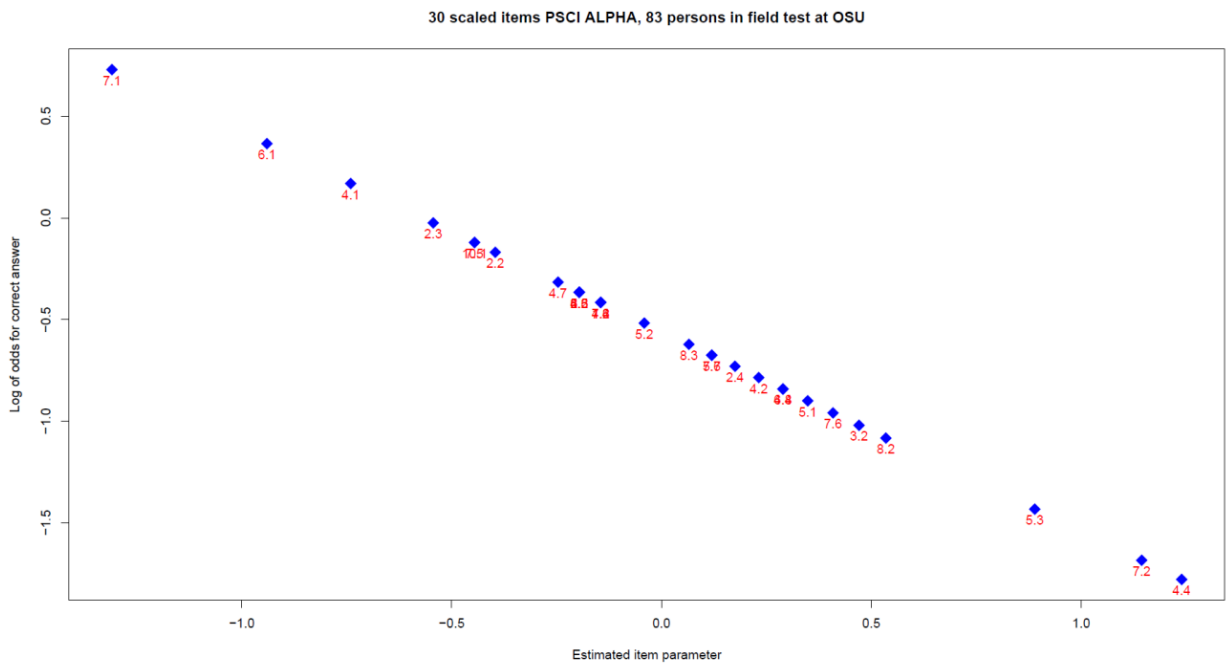

*Figure 1. The linear Rasch calibration of the items in PSCI-ALPHA independent on persons*

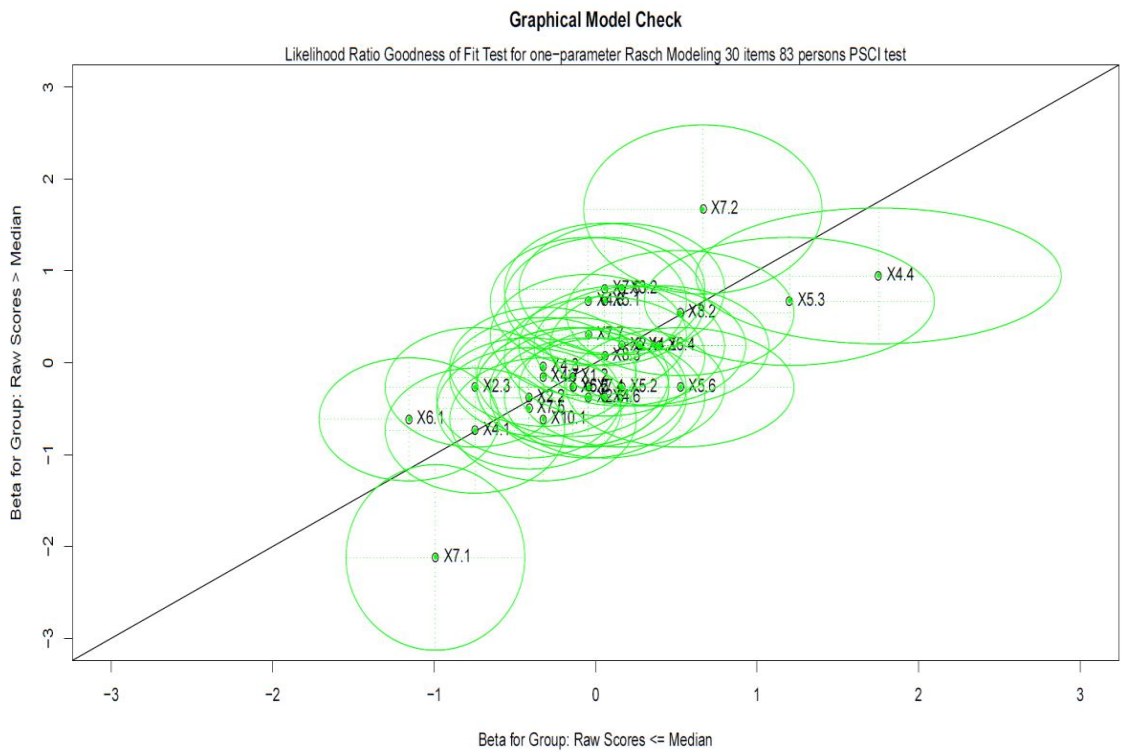

*Figure 2. LR goodness of fit median-split criterion PSCI-ALPHA. The beta-parameters are plotted for two groups to check scale invariance. The ellipses represent the 95% Confidence Interval for each item parameter.*

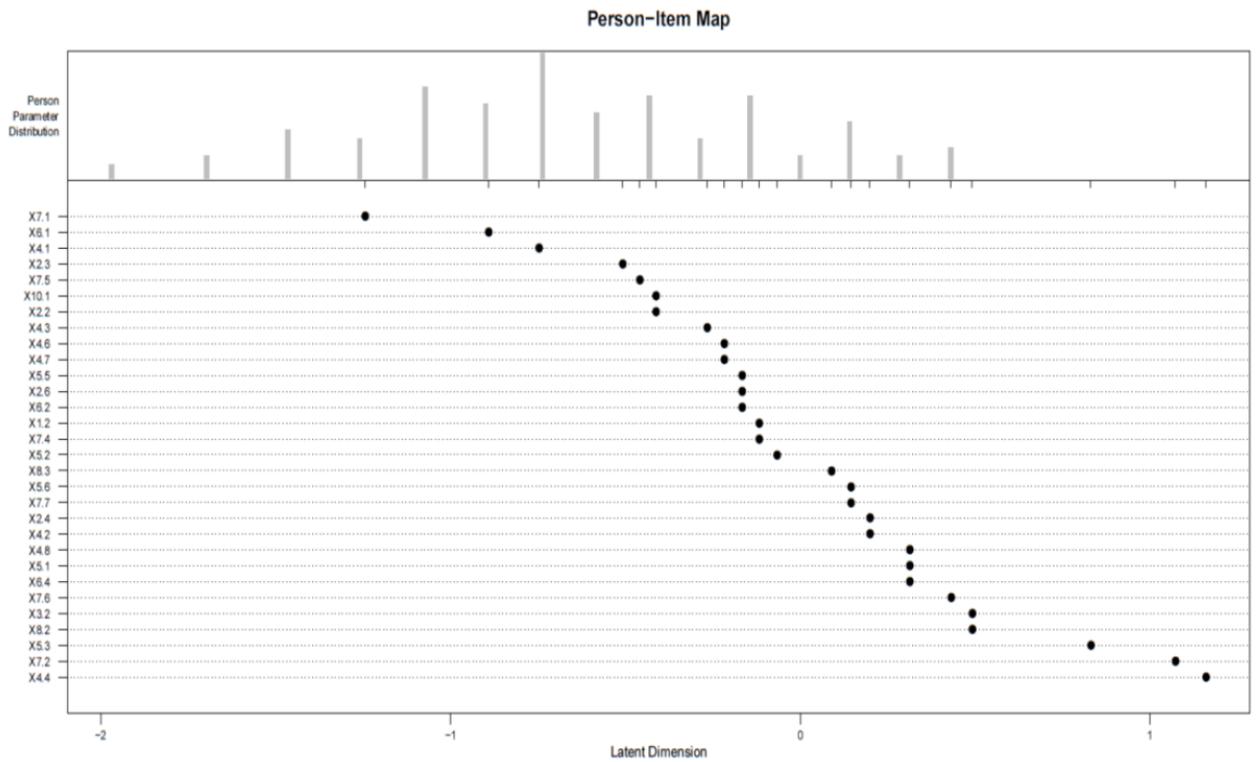

*Figure 3. Person/Item Map for PSCI-ALPHA 83 persons. The black dots represent items. The gray bars represent the distribution of the estimated person composite-ability.*

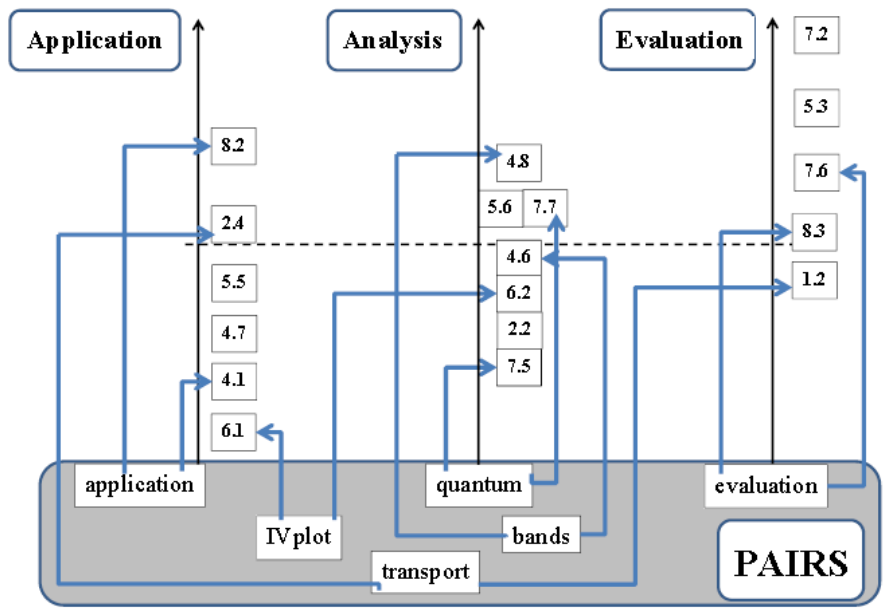

*Figure 4. Subsets in PSCI-BETA. The items in each subset are represented in the increasing order of the difficulty. The horizontal dashed line stands for the median difficulty. The subset Pairs links items from different subsets like the pairs "IV plot" and "transport" or items from the same subset.*

*Table 3. Test map for the PSC-BETA instrument*

| CAT. | CONCEPT | items # | items % | App. # | An. # | Ev. # |
|---|---|---|---|---|---|---|
| 1&2 | Transport | 3 | 17 | 1 | 1 | 1 |
| 4 | Bands | 4 | 22 | 2 | 2 | 0 |
| 5 | Carriers | 3 | 17 | 1 | 1 | 1 |
| 6 | I-V plot | 2 | 11 | 1 | 1 | 0 |
| 7 | Quantum | 4 | 22 | 0 | 2 | 2 |
| 8 | Statistics | 2 | 11 | 1 | 0 | 1 |
|  | TOTAL | 18 | 100 | 6 | 7 | 5 |

Notes. *CAT.=category; Und.=Understanding level; App.=Application level; An.=Analysis level; Ev.=Evaluation level*

### 4. Ability estimation based on a summary chart

The results of a measurement with PSCI-BETA on a 35-person independent sample are presented in the summary chart in **Figure 5**. The summary chart allows the instructor who applies the PSCI-BETA instrument to interpret the scores and make valid decisions. In this summary chart, column 1 gives the estimated person parameters for the 18-item set and columns 2 through 5 give the theta-parameters of each person for Application, Analysis, Evaluation, and Pairs. The color code helps to visualize the splitting of the persons in three categories. Yellow/grey stands for a ranking within a standard error from the median of the item-scale, implying a probability between 40% and 60% to solve that specific subtest. Green/dark stands for a ranking higher than a standard error from the median of the item-scale, while purple/light stands for a ranking lower than a standard error from the median. One may choose to use the more familiar letter grades for labeling these ability categories (i.e., C, B and above, and D and below.)

The color code in the summary chart allows a quick judgement about the general performance of the persons in the sample and about the consistency of the individual responses. These 35 persons have scored in average at the median of the PSCI-BETA item-scale, but lower than the median for the items in the subset Evaluation. Only person P13 has scored consistently above the median on all five scales. Person P3 has a low score for the subset Application which may denote missing knowledge, yet has fair ability in making inferences indicated by the high scores for Analysis and Evaluation." In contrast, person P10 could have achieved the high ability for the subset Evaluation by guessing because the rankings at the Application and Analysis subsets are under the median. Person P27 has a high ability for the subset Pairs, which denotes consistency in answering related items and thus a good quality of knowledge. The low ability for the subset Application may denote missing knowledge that P27 can remediate easily before the final exam for the course.

### 5. Learning gain estimation with a Rasch calibrated scale

The individual learning gain for a Rasch instrument is simply estimated by the change of the person parameter between pre- and post-test:

$$Ability\ Gain = \Delta\theta = \theta_{post} - \theta_{pre}$$

The Rasch metric for the learning gain is individual, specific, and objective. Average learning gains can be easily computed for comparing instruction effectiveness in different class settings. A learning gain measured by the PSCI-BETA between pre- and post-test can be considered evidence that instruction was focused on good quality conceptual knowledge. We found an average estimated learning gain of +0.8-logit on the 18-item scale for a 35-student PHYS3313 class at OSU. This value of the ability gain means that the odds to answer correctly rather than incorrectly to all the 18-items increased by a factor of 2.2.

The ability gain measured on the linear Rasch scale is non-linearly related to the increase in score; it represents the log of the odds ratio for answering correctly after instruction compared to before instruction. While popular logarithmic metrics are used for intensity of the sound and intensity of an earthquake, ability gain in an educational setting may be more difficult to interpret than the traditional metrics like Hake's gain[41].

Taking an example, assume a class of students with the average odds of 1:4 for answering correctly rather than incorrectly to the 18-item test before instruction. After an instruction with a +0.8-logit average ability gain the odds for this class at post-test become about 2:3. Or in probabilistic terms, if the average chance to answer correctly all the 18 items before instruction was 20%, after an instruction with a +0.8 average ability gain the average chance would become 26%. This represents a 6% increase in chance, which is not much. For another hypothetic group of students with the average chance of 43% for answering correctly all the 18 items at pre-test, a +.8-logit ability gain gives a 63% average chance to solve correctly all the 18 items at post-test. In the second fictitious case, the chance of solving correctly increased with 20%.

A popular measure of the average learning gain, the Hake's standardized gain[41], is related to the average mean (percentage) test grade $p$

$$g_{av.} = \frac{p_{POST} - p_{PRE}}{1 - p_{PRE}}$$

Assuming that the average chance to answer correctly to all the items in the test is approximately equal to the (percentage) mean grade $p$ for the test, we calculate Hake's gain for the two hypothetical examples discussed above. The first example of a +0.8-logit average ability gain was a class with a 20% average chance to solve the test correctly before instruction. Hake's gain for this class is

$$g_1 = \frac{0.26 - 0.20}{1 - 0.20} = 7.5\%$$

The second example of a +0.8-logit average ability gain was a class starting the instruction with the 43% average chance to solve the test correctly. For this second hypothetical situation Hake's gain is

$$g_2 = \frac{0.63 - 0.43}{1 - 0.43} = 35\%$$

The learning gain metrics based on probability favor the class with a higher average grade at pre-test. In the logarithmic metric of the average ability gain, the performances of the two fictitious classes are equal. Concluding, the ability gain metric measured on a Rasch scale has the advantage of objectively monitoring individual progress but could give a surprising message to those instructors who are usually reporting Hake's gain.

### C. IRT Modeling Results

The PSCI test was built to measure multiple facets of the conceptual knowledge of semiconductors. The items in each concept-category were analyzed separately using two-and three-parameter Item Response Theory (IRT) models. While the Rasch model considers all items equally discriminating and it is person distribution free, the IRT methods have the advantage to account for different discriminating power of the items and to give details about the persons' distribution. The Marginal Maximum Likelihood (MML) *ltm* package[42,43] available in the open source R language was employed for IRT modeling. The item parameter MML estimation depends on the person distribution, which may limit the generalizability to other samples of persons.

The fit of a 2PL IRT model[42] on each concept-category of items in the PSCI-ALPHA was interpreted as a confirmation that the items in the same category were addressing the same concept. The fit of the one-factor LTM[42] model on the entire set of items in PSCI-ALPHA was considered a piece of evidence that there is a common trait measured by all the 30 items. We reasonably labeled that common trait as "composite-ability to operate with conceptual knowledge of physics of semiconductors".

The results given by the 1PL IRT model in the *ltm* - MML package was compared to the Rasch model in the *eRm* -CML package by fitting the 18 items in the current version of the test PSCI-BETA. Both CML and MML methods ranked the 18 items identically, but the MML method inflated the distance between two adjacent items with a factor of 4.6 compared to CML. The average standard error was 8.1 times greater for MML than for CML.

| Person | 1<br>ppar.18 | 2<br>ppar.app | 3<br>ppar.ana | 4<br>ppar.eval | 5<br>ppar.pairs |
|---|---|---|---|---|---|
| P1 | -0.48 | 0.00 | -0.29 | -1.47 | -0.16 |
| P2 | -1.01 | 0.00 | -0.93 | -2.56 | -0.86 |
| P3 | 0.23 | -1.70 | 0.93 | 1.47 | 0.46 |
| P4 | -0.48 | 0.74 | -0.29 | -2.56 | -0.49 |
| P5 | -1.32 | -0.74 | -1.82 | -1.47 | -1.33 |
| P6 | -0.48 | -0.74 | -0.29 | -0.44 | -0.49 |
| P7 | -0.48 | 1.70 | -1.82 | -1.47 | -0.49 |
| P8 | -1.01 | -1.70 | -0.29 | -1.47 | -0.86 |
| P9 | -1.32 | -1.70 | -0.29 | -2.56 | -1.33 |
| P10 | -0.24 | -1.70 | -0.93 | 2.57 | -0.16 |
| P11 | 0.23 | 0.00 | 0.93 | -0.44 | 0.15 |
| P12 | -0.74 | 0.74 | -0.93 | -2.56 | -0.16 |
| P13 | 0.99 | 0.74 | 0.93 | 1.47 | 1.12 |
| P14 | -1.01 | -1.70 | -0.29 | -1.47 | -0.49 |
| P15 | 0.00 | -1.70 | 0.93 | 0.43 | 0.15 |
| P16 | -0.24 | 0.00 | -0.93 | 0.43 | -0.16 |
| P17 | -0.48 | -1.70 | 0.93 | -1.47 | -0.16 |
| P18 | -0.74 | 0.00 | -0.93 | -1.47 | -0.49 |
| P19 | -0.48 | 0.00 | -0.29 | -1.47 | 0.15 |
| P20 | -0.48 | 0.00 | -0.29 | -1.47 | -0.16 |
| P21 | 0.23 | 1.70 | 0.29 | -1.47 | 0.15 |
| P22 | -0.24 | 0.74 | -0.29 | -1.47 | -0.49 |
| P23 | 0.23 | 0.00 | 1.82 | -1.47 | 0.15 |
| P24 | -1.32 | -1.70 | -0.93 | -1.47 | -1.33 |
| P25 | -0.24 | -0.74 | 0.93 | -1.47 | -0.49 |
| P26 | 0.48 | 1.70 | -0.93 | 1.47 | 0.46 |
| P27 | 0.48 | -0.74 | 1.82 | 0.43 | 0.78 |
| P28 | -1.01 | -0.74 | -1.82 | -0.44 | -0.86 |
| P29 | 0.00 | -0.74 | 0.93 | -0.44 | 0.46 |
| P30 | -0.48 | 0.00 | -0.93 | -0.44 | -0.86 |
| P31 | -0.48 | -0.74 | 0.93 | -2.56 | -0.49 |
| P32 | 0.00 | 0.00 | 0.93 | -1.47 | -0.16 |
| P33 | -0.48 | -0.74 | -0.29 | -0.44 | 0.15 |
| P34 | 0.23 | 0.00 | 0.93 | -0.44 | 0.46 |
| P35 | -0.48 | 0.00 | -0.29 | -1.47 | -0.49 |

*Figure 5. Summary chart for PSCI-BETA administered in Spring2013 to a sample of 35 students at the end of the PHYS3313 course. The person parameters are calculated separately for each subset.*
*Note: ppar=person parameter*

## IV. Summary and conclusion

An assessment instrument for a specialty course was built with the collective contribution of instructors from nine universities. The relevancy and clarity of the initial pool of items was investigated by expert raters, while the interviews with the novices help identify language misrepresentations. Classical test analysis and IRT modeling were used for qualitative confirmation of the test reliability and construct validity. Although there were only 64 persons in field-testing who answered all the items in the PSCI-ALPHA, a Rasch Conditional Maximum Likelihood Calibration gave a person-invariant item scale. The item calibration, independent of the persons in the field test, was checked on an independent sample of students taking the same course.

The learning gain of each individual student is measured on the Rasch scale as the change of the parameter "person ability". The 18-item test PSCI-BETA allows the measurement of five different ability gains: of the basic knowledge of physics of semiconductors, of the knowledge applied at three different cognitive levels, and of the consistency of knowledge measured by correlated pairs of items. The independent subsets of items in PSCI-BETA facilitate a quick identification of the missing concepts or cognitive abilities. The partial score for the subset Pairs, modeled in the Partial Credit Rasch Model, accounts for the quality of knowledge. A color coded summary chart that converts the responses to person parameters allows a quick judgement about the general performance of the class of students and about the knowledge consistency of the individual persons.

Any team of instructors teaching a small size specialty course could follow the recipe described in this paper for calibrating a bank of items on a sample smaller than 100 students. Writing items dedicated to different levels in Bloom's taxonomy leads to subsets of calibrated Rasch scales. A summary chart tracks the incorrect answers to determine if they are due to the lack of knowledge at Remember/Understand levels or to the lack of training with multi-step higher level cognitive processes.

The objectivity of an RM instrument results from its independence from the person distribution in the specific sample used for item calibration. Objectivity can be attained only if data fit the RM model [13]. The PSCI-BETA instrument can be applied to other samples from the same population of persons who are enrolled or have been enrolled in an introductory course of semiconductor devices. Although the calibration of the items in PSCI-BETA is sample invariant, irregularities due to specific instructional strategies are not excluded.

The PSCI-BETA instrument, which is offered for use to all the instructors teaching introductory courses of physics of semiconductors, can be employed as a quick litmus test before and after instruction. The instructor interested to do this may contact the authors to receive a secure copy of the test and the score-to-chart conversion tool.

The authors intend to write and calibrate new items that would extend the range of measurement of the PSCI instrument. Having a larger data base of student answering patterns, it would be possible to assess the relationship between the ability gain and the instructional method for the physics of semiconductor courses like others have done employing the FCI[41] for Newtonian mechanics courses.

**ACKNOWLEDGEMENTS**

The authors would like to thank to Seth R. Bank from the University of Texas at Austin, Alan Cheville from the Bucknell University, Bruce van Dover from the Cornell University, Stephen Hersee from the University of New Mexico, Patrick McCann from the Oklahoma University , Jamie Phillips from the University of Michigan, Sandra Selmic from the Louisiana Tech University, Daryoosh Vashaee from the North Carolina State University, and Robert Hauenstein, Jerzy Krasinski, Stephen McKeever, and James P. Wicksted all three from the Oklahoma State University for listing the top ten concepts for the physics of semiconductors. The authors are especially grateful to Alan Cheville, Seth Bank, Robert Hauenstein, and Jerzy Krasinski for their contribution to the writing of the initial pool of items, and to Jeremy Penn from the Office of Assessment at the North Dakota State University for the discussions on the statistical modeling. Special thanks to Mario Borunda, Alexander Khanov, and Flera Rizatdinova from Oklahoma State University for applying the PSCI test to their students.

**APPENDIX**

Text books surveyed for the top ten concepts for the physics of semiconductors

| Authors |
|---|
| Anderson & Anderson [44] |
| Boylestad & Nashelsky [45] |
| Jaeger [46] |
| Neamen [47] |
| Streetman & Banerjee [48] |
| Van Zeghbroeck [49] |
| Pierret [50] |
| Sze [51] |
| Carroll[52] |